\pdfoutput=1

\documentclass[11pt]{article}

\usepackage[final]{acl}

\usepackage{times}
\usepackage{latexsym}

\usepackage[T1]{fontenc}

\usepackage[utf8]{inputenc}

\usepackage{microtype}

\usepackage{inconsolata}

\usepackage{hyperref}
\usepackage{url}
\usepackage{latexsym}
\usepackage{xcolor}
\usepackage{amssymb}
\usepackage{pifont}
\usepackage{graphicx}
\usepackage{booktabs}
\usepackage{makecell}
\usepackage{appendix}
\usepackage{enumitem}
\usepackage{comment}
\usepackage{multirow}
\usepackage{textcomp}
\usepackage{amsmath}

\newcommand{\CHECK}{\textcolor{green}{\ding{51}}} 
\newcommand{\CROSS}{\textcolor{red}{\ding{55}}} 

\title{LauraGPT: Listen, Attend, Understand, and Regenerate Audio with GPT}

\author{Zhihao Du$^*$, Jiaming Wang$^*$, Qian Chen, Jin Xu, Zhifu Gao, Zerui Li \\ {\bf Kai Hu, Xiaohuan Zhou, Ziyang Ma, Yunfei Chu, Wen Wang} \\ {\bf Siqi Zheng, Chang Zhou, Zhijie Yan, Shiliang Zhang$^\dagger$}\thanks{equal contribution. $^\dagger$ correspondence author.} \\
Damo Academy, Alibaba Group, China\\
 \texttt{\{neo.dzh,sly.zsl\}@alibaba-inc.com}}

\begin{document}
\maketitle
\begin{abstract}
Generative Pre-trained Transformer~(GPT) models have achieved remarkable performance on various natural language processing tasks, and have shown great potential as backbones for audio-and-text large language models (LLMs). Previous mainstream audio-and-text LLMs use discrete audio tokens to represent both input and output audio; however, they suffer from performance degradation on tasks such as automatic speech recognition, speech-to-text translation, and speech enhancement over models using continuous speech features. In this paper, we propose \textbf{LauraGPT}, a novel unified audio-and-text GPT-based LLM for audio recognition, understanding, and generation. LauraGPT is a versatile LLM that can process both audio and text inputs and generate outputs in either modalities. We propose a novel data representation that combines continuous and discrete features for audio: LauraGPT encodes input audio into continuous representations using an audio encoder and generates output audio from discrete codec codes. We propose a one-step codec vocoder to overcome the prediction challenge caused by the multimodal distribution of codec tokens. We fine-tune LauraGPT using supervised multi-task learning. Extensive experiments show that LauraGPT consistently achieves comparable to superior performance compared to strong baselines on a wide range of audio tasks related to content, semantics, paralinguistics, and audio-signal analysis, such as automatic speech recognition, speech-to-text translation, text-to-speech synthesis, speech enhancement, automated audio captioning, speech emotion recognition, and spoken language understanding. 
\end{abstract}

\section{Introduction}
\label{sec:intro}
Large language models (LLMs) are neural networks that generate natural language texts based on a given context. 
LLMs can learn from massive amounts of text data and mimic human language to acquire human knowledge. LLMs such as GPT-4~\citep{DBLP:journals/corr/abs-2303-08774}, PaLM2~\citep{DBLP:journals/corr/abs-2305-10403}, LLaMA~\citep{DBLP:journals/corr/abs-2302-13971} have demonstrated impressive capabilities across various domains, exhibiting zero-shot generalization without the need for task-specific fine-tuning. However, these models are primarily limited to processing text data. 

Recent research aims to seamlessly integrate text and audio since they are two important modalities for human communication. These efforts include \textbf{Audio-to-Text LLMs}~\citep{DBLP:journals/corr/abs-2212-04356,DBLP:journals/corr/abs-2303-01037,DBLP:journals/corr/abs-2305-11834,DBLP:journals/corr/abs-2310-02973,DBLP:journals/corr/abs-2310-13289,DBLP:journals/corr/abs-2311-07919}, which can convert audio input into text and perform tasks such as automatic speech recognition (ASR) and spoken language understanding (SLU); \textbf{Text-to-Audio LLMs}~\citep{DBLP:journals/corr/abs-2310-00704,DBLP:journals/corr/abs-2312-15821,DBLP:conf/iclr/KreukSPSDCPTA23,DBLP:journals/corr/abs-2308-05734,DBLP:conf/icml/HuangHY0LLYLYZ23,DBLP:journals/corr/abs-2301-02111}, which can convert text input into audio and perform tasks such as text-to-speech synthesis (TTS) and text-to-music synthesis. An emerging line of research focuses on develop more universal and comprehensive \textbf{Audio-and-Text LLMs}~\citep{DBLP:conf/acl/AoWZ0RW0KLZWQ0W22,DBLP:journals/corr/abs-2105-03070,DBLP:journals/corr/abs-2305-11000,DBLP:journals/corr/abs-2305-16107,DBLP:journals/corr/abs-2306-12925,DBLP:journals/corr/abs-2304-12995}, which can support audio-and-text tasks, that is, process and generate both audio and text and perform tasks such as speech enhancement (SE) and speech-to-speech translation (S2ST), in addition to tasks supported by audio-to-text and text-to-audio LLMs. Audio-to-text and text-to-audio LLMs can be considered as subsets of audio-and-text LLMs. 

Audio-and-Text LLMs can be categorized into two directions. One direction builds \textbf{a collaborative AI system} using LLMs as controllers to interface specialized audio models, such as ASR and TTS models, to support various audio-and-text tasks~\citep{DBLP:journals/corr/abs-2303-17580,DBLP:journals/corr/abs-2304-12995}.  These methods have serious drawbacks, including high complexity, significant resource consumption, and unavoidable error accumulation problems. The other direction develops a \textbf{unified Audio-and-Text LLM} leveraging LLMs as the backbone to support audio-and-text tasks~\citep{DBLP:conf/acl/AoWZ0RW0KLZWQ0W22,DBLP:journals/corr/abs-2105-03070,DBLP:journals/corr/abs-2305-16107,DBLP:journals/corr/abs-2306-12925}. Decoder-only audio-and-text LLMs~\citep{DBLP:journals/corr/abs-2305-11000,DBLP:journals/corr/abs-2305-16107,DBLP:journals/corr/abs-2306-12925} are the dominant technique under this category. These models convert continuous audio into discrete tokens and integrate text and audio tokens into unified vocabulary. These models suffer from information loss from quantization of speech signals into discrete tokens, which leads to notable performance degradation on ASR compared to models using continuous speech features~\citep{DBLP:journals/corr/abs-2311-04534,DBLP:journals/corr/abs-2305-18108,DBLP:journals/corr/abs-2309-07377,DBLP:journals/corr/abs-2309-10922}. In this paper, we focus on improving the second category of unified Audio-and-Text LLMs. Moreover, recent advances in audio generation from unified audio-and-text LLMs~\citep{DBLP:journals/corr/abs-2301-02111,DBLP:journals/corr/abs-2305-16107} discretize speech into codec codes, then use an autoregressive language model (LM) to predict output tokens from the first quantizer and use a non-autoregressive model to predict tokens from the other quantizers individually. One limitation of this mechanism is that it needs many prediction steps (hence called \textbf{multi-step audio synthesis scheme}) to generate good quality speech. Another limitation is that predicting the indices of the other codec groups is challenging due to the multi-modal distribution nature of codec tokens~\citep{DBLP:journals/icassp/lmcodec}.

To overcome the drawbacks of existing \textit{unified audio-and-text LLMs}, we propose \textbf{LauraGPT}, a novel \textbf{unified Audio-and-Text LLM} based on the GPT framework for audio recognition, understanding, and generation. LauraGPT is a versatile LLM that can process both audio and text inputs and generate outputs in either modalities, with a single model. We propose \textbf{a novel data representation that combines continuous and discrete features for audio}: LauraGPT encodes input audio into continuous representations using an audio encoder and generates output audio from discrete codec codes. This data representation improves the performance of audio-input tasks and also facilitates joint autoregressive modeling of audio and text features for audio generation tasks. 

 We also propose \textbf{a one-step codec vocoder in LauraGPT to address the two limitations of the popular multi-step audio synthesis scheme}. Our one-step codec vocoder uses a transformer-based predictor to estimate the sum of all codec token groups instead of the individual indices, by minimizing the reconstruction losses. Our approach simplifies the audio generation process to a \textit{single} feed-forward calculation and also overcomes the prediction challenge caused by the multi-modal distribution of codec tokens.
 
We fine-tune LauraGPT using \textbf{supervised multi-task learning on diverse audio tasks}, including tasks focusing on content, semantics, paralinguistics, and audio-signal analysis, such as ASR, speech-to-text translation (S2TT), TTS, SE, automated audio captioning (AAC), speech emotion recognition (SER), and SLU. \textbf{Comprehensive experiments show that, to the best of our knowledge, LauraGPT\footnote{Demos  are available at \url{https://lauragpt.github.io}} consistently achieves comparable to superior performance compared to strong baselines on the largest and the most diverse set of audio recognition, understanding, and generation tasks among existing decoder-only unified audio-and-text LLMs focusing on these tasks}~\citep{DBLP:journals/corr/abs-2305-11000,DBLP:journals/corr/abs-2305-16107,DBLP:journals/corr/abs-2306-12925}. The results are remarkable since existing general speech models either focus solely on speech recognition and understanding tasks but neglect speech generative tasks, or support speech generation but suffer from severe performance degradation on speech recognition and understanding tasks.
\section{Related Work}
\label{sec:related_work}
\noindent \textbf{Audio-to-Text LLMs}
Audio-to-Text LLMs can generate text from audio inputs. Whisper~\citep{DBLP:journals/corr/abs-2212-04356}  and USM~\citep{DBLP:journals/corr/abs-2303-01037} can perform speech recognition and translation across multiple languages and domains. Pengi~\citep{DBLP:journals/corr/abs-2305-11834} is an audio LM that formulates audio tasks as text-generation tasks. UniverSLU~\citep{DBLP:journals/corr/abs-2310-02973} is a universal SLU model that supports various speech classification and sequence generation tasks. SALMONN~\citep{DBLP:journals/corr/abs-2310-13289} and Qwen-Audio~\citep{DBLP:journals/corr/abs-2311-07919} integrate pre-trained text LLMs with separate speech and audio encoders into a single multimodal model. 

\noindent \textbf{Text-to-Audio LLMs}
Text-to-Audio LLMs can convert text input into audio output and perform tasks such as TTS or text-to-music synthesis.
Recently, two prominent categories of approaches have emerged for generating audio from text prompts.
In the first category, continuous representations such as utterance-level embeddings~\citep{DBLP:journals/corr/abs-2206-04769, DBLP:conf/icml/LiuCYMLM0P23, DBLP:conf/icml/HuangHY0LLYLYZ23} and Mel-frequency spectrograms~\citep{DBLP:journals/corr/abs-2305-15255} are used as the targets. 
However, continuous representations present a challenge for unified modeling of text and audio within a single LM.
In the second category, discrete codec tokens are employed as audio representations and generated by diffusion models~\citep{DBLP:journals/taslp/YangYWWWZY23} or autoregressive LMs~\citep{DBLP:conf/iclr/KreukSPSDCPTA23,DBLP:journals/taslp/BorsosMVKPSRTGTZ23,DBLP:journals/corr/abs-2306-05284,DBLP:journals/corr/abs-2301-02111}. 
Among models in the second category, in models such as AudioGen~\citep{DBLP:conf/iclr/KreukSPSDCPTA23}, AudioLM~\citep{DBLP:journals/taslp/BorsosMVKPSRTGTZ23}, and MusicGen~\citep{DBLP:journals/corr/abs-2306-05284}, multiple output heads are used after the LM to predict synchronized or delayed groups of codec tokens. 
However, this mechanism is only suitable for audio generation and may not be applicable to diverse audio-and-text tasks. 
Alternatively, in VALL-E~\citep{DBLP:journals/corr/abs-2301-02111}, the LM predicts output tokens of the first quantizer, while tokens of the remaining quantizers are predicted by a non-autoregressive model one by one.
This mechanism requires numerous prediction procedures to generate acceptable speech quality. Moreover, the indices of 
the remaining codec groups are challenging to predict due to the multi-modal distribution nature of codec tokens~\citep{DBLP:journals/icassp/lmcodec}.

\noindent \textbf{Audio-and-Text LLMs}
Audio-and-Text LLMs can process and generate both audio and text, which can be categorized into two directions. One direction uses LLMs as controllers to interface specialized audio models, such as ASR and TTS models, to enable direct audio interaction with LLMs and support various audio-and-text tasks, such as HuggingGPT~\citep{DBLP:journals/corr/abs-2303-17580} and AudioGPT~\citep{DBLP:journals/corr/abs-2304-12995}. However, these models are complex, resource-intensive, and prone to error accumulation. The second direction uses LLMs as the backbone for a unified model that handles audio-and-text tasks~\citep{DBLP:conf/acl/AoWZ0RW0KLZWQ0W22,DBLP:journals/corr/abs-2105-03070,DBLP:journals/corr/abs-2305-16107,DBLP:journals/corr/abs-2306-12925}. SpeechT5~\citep{DBLP:conf/acl/AoWZ0RW0KLZWQ0W22} and SpeechNet~\citep{DBLP:journals/corr/abs-2105-03070} perform various speech tasks with an encoder-decoder model, but they require modal-specific pre-nets and post-nets to deal with different input\&output modalities. VioLA~\citep{DBLP:journals/corr/abs-2305-16107}, AudioPaLM~\citep{DBLP:journals/corr/abs-2306-12925}, SpeechGPT~\citep{DBLP:journals/corr/abs-2305-11000}, and SpeechGen~\citep{DBLP:journals/corr/abs-2306-02207} use decoder-only Transformers to model discrete audio tokens and text tokens as a shared vocabulary, but they suffer from information loss from quantization of audio signals into discrete tokens~\citep{DBLP:journals/corr/abs-2311-04534,DBLP:journals/corr/abs-2305-18108,DBLP:journals/corr/abs-2309-07377,DBLP:journals/corr/abs-2309-10922}.



\begin{figure*}[htb]
  \centering
  \includegraphics[width=0.9\linewidth]{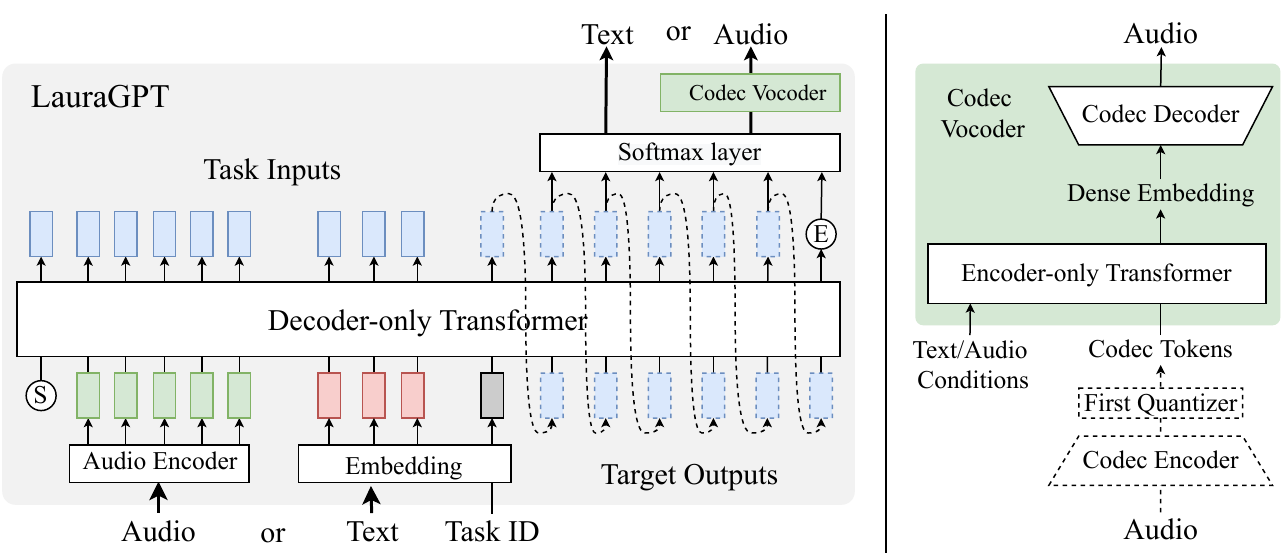}
  \caption{The overview of the proposed LauraGPT model. The right part provides an enlarged view of the one-step Codec Vocoder (Section~\ref{sec:codec_vocoder}) in LauraGPT. The dashed modules are only used in the training stage. \textcircled{S} and \textcircled{E} denote the ``start of sequence'' and ``end of sequence'' tokens. We omit the text tokenizer and detokenizer for simplicity.}
  \label{fig:overall}
  \vspace{-0.2cm}
\end{figure*}

\vspace{-2mm}
\section{Methodology}
\vspace{-1mm}
\label{sec:methods}
Figure~\ref{fig:overall} depicts the architecture of the proposed LauraGPT.
Section \ref{sec:gpt-backbone} describes the audio encoder, the text tokenizer, and the modified GPT LM for unified audio-and-text modeling. 
Section~\ref{sec:audio-tokenizer} elaborates the audio tokenizer.
Section~\ref{sec:codec_vocoder} introduces an efficient one-step codec vocoder for converting audio tokens into high-quality raw waveforms.
Section~\ref{sec:task-details} describes the multi-task fine-tuning and shows that LauraGPT provides an extensible framework for supporting more complex tasks.

\vspace{-1mm}
\subsection{Modified Language Model for Unifying Audio-and-Text Modeling}
\label{sec:gpt-backbone}
For audio inputs, different from other audio-and-text LLMs using discrete tokens to represent audio inputs, we extract the log-compressed Mel spectrogram features and convert them into  \textit{continuous representations} using a Conformer-based audio encoder.
Text inputs and outputs are tokenized using the Qwen tokenizer \cite{qwen}, which inherits the tiktoken tokenizer~\citep{tiktoken} and incorporates additional augmentations for commonly used characters and words in different languages.
The tokenized input text undergoes embedding matrix transformation to generate dense vectors.
The audio representations and text embeddings have the same dimension $D$.
The Conformer-based encoder is initialized with weights from a pre-trained ASR model~\citep{gao2023funasr}. 
Since batch normalization can lead to endless loop decoding, we replace it with layer normalization in the Conformer-based encoder (details are in Appendix~\ref{sec:normlization}).

To achieve audio generation capabilities, the audio outputs are discretized into tokens using an audio tokenizer (Section~\ref{sec:audio-tokenizer}) to obtain \textit{discrete representations} and the softmax output layer is augmented with the audio tokens. 
As a result, the weight matrix $\mathbf{W}$ in the output layer is of size $(N+M+L) \times D$ and is utilized to calculate the logits for audio and text tokens at each position, where $N$, $M$, and $L$ denote the vocabulary sizes of text, audio, and task tokens, respectively. Task tokens are used to inform the model which task should be performed. Note that in order to control the sequence length, we perform the low frame rate~(LFR) method~\citep{gao2020san-m} to downsample audio inputs to 60ms and only select the first codec group of the audio outputs.

Based on the aforementioned representations, the GPT backbone is trained to model various audio and text tasks by minimizing the cross-entropy loss:
\vspace{-6mm}
\begin{equation}
\label{eq:LM_loss}
    \mathcal{L}_{LM}=-\frac{1}{T_v}\sum_{j=1}^{T_v}{
    \log p_\theta\left(
    \mathbf{v}_j | \mathbf{u}_{1:T_u}, \mathbf{u}_{task},\mathbf{v}_{1:j-1}
    \right)
    }
\end{equation}
where $\mathbf{u}$ denotes the input embeddings with a sequence length $T_u$ and 
$\mathbf{v}$ represents the sequence of target tokens with a length $T_v$. 
To specify a task, a special task-related token $\mathbf{u}_{task}$ is inserted between the input embeddings and output tokens.
Note that only the losses of outputs are taken into account, while losses on inputs and task token embeddings are masked out.
After the final output layer, 
audio tokens are decoded to raw waveforms using a codec vocoder (Section \ref{sec:codec_vocoder}).
Since it is challenging to train an LLM from scratch with limited data and computational resources, we use the open-source GPT LLM, Qwen~\citep{qwen}, as the backbone. Qwen is pre-trained on a diverse corpus covering various domains in English and Chinese and supports 8192 context length. Compared with other open-source GPT models with similar model sizes, Qwen models demonstrate impressive competitiveness, achieving better performance on widely used benchmarks, especially on Chinese tasks~\citep{qwen}.
Within LauraGPT, all parameters including the Qwen backbone are jointly optimized, except for the codec vocoder, which is trained independently and kept frozen during both training and inference stages of LauraGPT.

\vspace{-2mm}
\subsection{Audio Tokenizer}
\vspace{-1mm}
\label{sec:audio-tokenizer}
For audio generation, we utilize a codec model as the audio tokenizer to extract \textit{discrete} representations.
Our codec model shares a similar architecture as EnCodec~\citep{defossez2022highfi}, which comprises convolutional recurrent encoder and decoder~\citep{DBLP:conf/interspeech/TagliasacchiLMR20} and a residual vector quantizer (RVQ)~\citep{vasuki2006review}.
We enhance the original EnCodec model with the following modifications:
1) Add reconstruction losses in the magnitude spectrum domain to improve the quality of middle- and high-frequency signals.
2) Stack five strided convolution blocks with strides of $[8, 5, 4, 2, 2]$ to address the challenge of long sequence lengths, resulting in a token rate of 25Hz for each token group. 
3) Use 32 quantizers with structured dropout in the RVQ module, each with vocabulary size 1024. This revision improves speech quality with more quantizers while preserving most information in the shallow quantizers.
The encoder and the \textit{first RVQ quantizer} are used as the audio tokenizer, and \textbf{the outputs of the first quantizer are used as the audio tokens}. The choice of the first $N$ RVQ quantizers to use is a tradeoff between performance and sequence length (hence efficiency). The remaining quantizers and the decoder are only used when training the codec model. Details of training and the pre-trained codec model are in ~\cite{du2023funcodec}. 



\vspace{-3mm}
\subsection{One-step Codec Vocoder for Audio Generation}
\label{sec:codec_vocoder}
We propose a one-step codec vocoder in LauraGPT to generate waveforms from the audio tokens, which are extracted from the \textit{first} quantizer as described in Section~\ref{sec:audio-tokenizer}. Our vocoder comprises two components: a transformer-based predictor and a codec decoder.
The predictor is trained to estimate the summation of codec embeddings from the 32 RVQ quantizers by minimizing the L1 and L2 distances between the predicted embeddings $\hat{\mathbf{E}}$ and their corresponding ground truth $\mathbf{E}$:
\vspace{-0.2cm}
\begin{equation}
\mathcal{L}_{pre}=\sum_{t,i}^{T,D_c}{|\mathbf{E}_{t,i}-\hat{\mathbf{E}}_{t,i}|_1 + |\mathbf{E}_{t,i}-\hat{\mathbf{E}}_{t,i}|_2}
\vspace{-0.1cm}
\end{equation}
where $T$ denotes the total number of frames and $D_{c}$ denotes the dimension of the codec embeddings. 
After obtaining the estimated embeddings, the decoder of an pre-trained codec model is utilized to reconstruct the raw audio waveforms. 

\textbf{Alongside the predicted audio tokens from the LLM, text and audio inputs are used as conditions and fed to the predictor}.
For zero-shot TTS task, the text inputs serve as a condition as well as the prompt audio features. For SE task, the input noisy speech features are employed as conditions.
Such text and audio conditionings allow the model to generate high-quality audio signals by leveraging the diverse information in prompt audios and noisy speeches, which is lacked in the discrete tokens (output from the first quantizer). Therefore, different from existing Text-to-Audio LLMs, \textbf{our approach simplifies the audio generation process to a single feed-forward calculation and overcomes the prediction challenge caused by the multi-modal distribution of codec tokens}.

\vspace{-0.1cm}
\subsection{Multi-task Finetuning}
\label{sec:task-details}
\vspace{-0.1cm}
\paragraph{Basic Tasks}
We unify modeling of the following \textit{basic tasks} in the single LauraGPT model and use these tasks for multi-task fine-tuning:
Automatic Speech Recognition (\textbf{ASR}), Spoken Language
Understanding (\textbf{SLU}), Speech-to-Text Translation (\textbf{S2TT}), Speech Emotion Recognition (\textbf{SER}), Automated Audio Captioning (\textbf{AAC}), Speech Enhancement (\textbf{SE}), and Text-to-speech Synthesis (\textbf{TTS}). Task definitions are in Appendix~\ref{sec:task-intro}.

\vspace{-0.2cm}
\paragraph{Unified Task Expression}
\label{sec:task-formulation}
LauraGPT operates based on a unified task expression: \texttt{[input embeddings, task ID, output tokens]}.
With the same inputs, the desired outputs can differ across tasks. For instance, ASR and S2TT tasks require different outputs even for the same audio input.
Task tokens are included in both input embedding and output weight matrices.
The TTS task takes text embeddings as inputs, while the ASR, S2TT, SLU, SE, ACC, and SER tasks take audio encodings as inputs. 
The TTS and SE tasks use audio tokens as the target outputs, while the remaining tasks use text tokens as the target outputs.

\vspace{-0.2cm}
\paragraph{Support More Complex Tasks}
\label{sec:task-combination}
With its modularized design, LauraGPT provides an extensible framework to support complex tasks. By breaking a task into sub-tasks among the basic tasks and cascading the raw inputs and model outputs of sub-tasks, LauraGPT can perform more complex tasks.
For example, we demonstrate that LauraGPT is capable of performing the advanced speech-to-speech translation (S2ST) task by combining the S2TT and TTS tasks.
Initially, a sequence is constructed to translate the speech content into the target language text using the S2TT task token: \texttt{[audio encoding, <S2TT>]}. 
Subsequently, the translated text is combined with the TTS task token to synthesize speech: \texttt{[text embedding, <TTS>]}. 
If maintaining the speaker identity is desired, the original inputs and content can be incorporated to perform \textit{personalized TTS}. This can be achieved with an input sequence as \texttt{[ASR recognized text embedding, S2TT translated text embedding, <TTS>, audio token of input speech]}, where \texttt{ASR recognized text embedding} is obtained using the ASR task: \texttt{[audio encoding, <ASR>]}. This approach treats the bilingual text as the complete input and allows the model to generate an output sequence of codec tokens while maintaining the same speaker identity. Audio samples of S2ST can be found on the demo site. More examples of complex tasks are in Appendix~\ref{appendix:more-complex-tasks}.


\vspace{-2mm}
\section{Experimental Settings}
\label{sec:expriment}
\vspace{-1mm}
\noindent \textbf{Model Architecture} The Conformer-based audio encoder consists of 32 conformer blocks.
Each block consists of a feed-forward module with 1536 units, an attention module with 16 heads and a dimension of 512, a convolutional module including the pointwise and depthwise convolution layers, and a second feed-forward module with 1536 units. 
Sinusoidal positional encoding is applied on the audio inputs. 
For a trade-off between performance and training efficiency, we use Qwen-1.8B\footnote{\url{https://github.com/QwenLM/Qwen}} as the backbone and LauraGPT has 2B parameters.
Qwen-1.8B comprises 24 transformer layers with a hidden size 2048 and 16 attention heads. 
\textbf{Although Conformer and Qwen-1.8B are selected as the audio encoder and GPT backbone, they can be replaced by other encoders and GPT models}.

\noindent \textbf{Training Setup} In all experiments, we initialize the Qwen backbone and audio encoder with the pre-trained checkpoints. We then optimize the model parameters through multi-task fine-tuning. The training\&test datasets and evaluation metrics are presented in Appendix~\ref{sec:training-datasets} and ~\ref{sec:evaldata-metrics}. Appendix~\ref{sec:detail_training_setup} describes the three-stage training process to address the significant variation in data volume across different tasks, and details the inference process.

\begin{table*}[htb]
\setlength\tabcolsep{3pt}
\centering 
\caption{Results from the \textbf{SOTA}, a \textit{comparable} baseline, and our \textbf{LauraGPT}, in that order, on \textbf{speech recognition, understanding, and generation tasks}.
The better results between the baseline and LauraGPT are in bold.}
\label{tab:overall_res}
\scalebox{0.79}{
\begin{tabular}{l c c c c} 
\toprule 
\textbf{Task}    & \textbf{Test Set}  & \textbf{Metric}  & \textbf{Model}  & \textbf{Performance} \\
\midrule
\multirow{12}{*}{\textbf{ASR}}   & \multirow{3}{*}{AISHELL-1 test} & \multirow{3}{*}{CER~$\downarrow$} & Qwen-Audio~\citep{DBLP:journals/corr/abs-2311-07919} & 1.3  \\
& & & MMSpeech-large~\citep{zhou2022mmspeech} & 1.9 \\
& & & LauraGPT & \textbf{1.8} \\
\cmidrule(lr){2-5}
& \multirow{3}{*}{AISHELL-2 test-ios} & \multirow{3}{*}{CER~$\downarrow$} & Paraformer-large~\citep{gao2023funasr} & 2.9  \\
& & & MMSpeech-large~\citep{zhou2022mmspeech} & 3.9 \\
& & & LauraGPT & \textbf{3.2} \\
\cmidrule(lr){2-5}
& \multirow{3}{*}{LibriSpeech test-clean} & \multirow{3}{*}{WER~$\downarrow$} & Qwen-Audio~\citep{DBLP:journals/corr/abs-2311-07919} & 2.0  \\
& & & Whisper Large V2~\citep{radford2023robust} & \textbf{2.5} \\
& & & LauraGPT & 4.4 \\
\cmidrule(lr){2-5}
& \multirow{3}{*}{LibriSpeech test-other} & \multirow{3}{*}{WER~$\downarrow$} & Qwen-Audio~\citep{DBLP:journals/corr/abs-2311-07919} & 4.2  \\
& & & Whisper Large V2~\citep{radford2023robust} & \textbf{4.9} \\
& & & LauraGPT & 7.7 \\
\midrule
\multirow{3}{*}{\textbf{SLU}}   & \multirow{3}{*}{SLURP test} & \multirow{3}{*}{Intent ACC~$\uparrow$~|~SLU-F1~$\uparrow$ } & UniverSLU~\citep{DBLP:journals/corr/abs-2310-02973} & 90.5~|~80.5  \\
& & & Wav2Vec 2.0~\citep{DBLP:journals/corr/abs-2106-04624} & 85.3~|~\textbf{74.6} \\
& & & LauraGPT & \textbf{87.9} | 73.5 \\
\midrule                      
\multirow{6}{*}{\textbf{S2TT}}   & \multirow{3}{*}{\makecell{BSTC dev\\(Zh$\rightarrow$EN)}} & \multirow{3}{*}{BLEU~$\uparrow$} & - & -  \\
& & & Cascade-System~\cite{dataset_bstc} & \textbf{18.2} \\
& & & LauraGPT & 17.8 \\
\cmidrule(lr){2-5}
& \multirow{3}{*}{\makecell{CoVOST2 test set\\(En$\rightarrow$Zh)}} & \multirow{3}{*}{BLEU~$\uparrow$} & Qwen-Audio~\citep{DBLP:journals/corr/abs-2311-07919} & 41.5  \\
& & & EncDec-Attn~\cite{dataset_covost_2} & 25.4 \\
& & & LauraGPT & \textbf{38.5} \\
\midrule
\multirow{3}{*}{\textbf{SER}}   & \multirow{3}{*}{MELD test} & \multirow{3}{*}{WA~$\uparrow$~|~UA~$\uparrow$~|~WF1~$\uparrow$ } & Qwen-Audio~\citep{DBLP:journals/corr/abs-2311-07919} & 0.557~|~-~|~-  \\
& & & Vesper-12~\citep{chen2023vesper} & \textbf{0.535}~|~0.268~|~0.480 \\
& & & LauraGPT & 0.507~|~\textbf{0.312}~|~\textbf{0.492} \\
\midrule
\multirow{3}{*}{\textbf{AAC}}   & \multirow{3}{*}{Clotho eval} & \multirow{3}{*}{SPICE~$\uparrow$~|~CIDEr~$\uparrow$~|~SPIDEr~$\uparrow$ } & Qwen-Audio~\citep{DBLP:journals/corr/abs-2311-07919} & 0.14~|~0.44~|~0.29  \\
& & & Ensemble~\citep{koizumi2020ntt} & \textbf{0.09}~|~\textbf{0.32}~|~\textbf{0.21} \\
& & & LauraGPT & 0.08~|~0.22~|~0.15 \\
\midrule
\multirow{3}{*}{\textbf{SE}}   & \multirow{3}{*}{\makecell{Mixup of LibriSpeech\\test-clean, FSD50K and\\noise-92}} & \multirow{3}{*}{\makecell{WER~$\downarrow$~|\\PESQ~$\uparrow$~|~STOI~$\uparrow$}} & - & - \\
& & & CMGAN~\citep{cao22_interspeech} & \textbf{12.29}~|~2.95~|~\textbf{91.0}  \\
& & & LauraGPT & 15.94~|~\textbf{2.97}~|~88.0 \\
\midrule
\multirow{6}{*}{\textbf{TTS}}   & \multirow{3}{*}{AISHELL-1} & \multirow{3}{*}{\makecell{CER~$\downarrow$~|\\SECS~$\uparrow$~|~MOSNet~$\uparrow$}} & VALL-E Phone~\citep{DBLP:journals/corr/abs-2301-02111} & 4.75~|~0.91~|~3.22  \\
 & & & VALL-E Token~\citep{DBLP:journals/corr/abs-2301-02111} & \textbf{6.52}~|~\textbf{0.91}~|~\textbf{3.19} \\
& & & LauraGPT & 6.91~|~0.90~|~3.14 \\
\cmidrule(lr){2-5}
& \multirow{3}{*}{LibriTTS} & \multirow{3}{*}{\makecell{WER~$\downarrow$~|\\SECS~$\uparrow$~|~MOSNet~$\uparrow$}} & VALL-E Phone~\citep{DBLP:journals/corr/abs-2301-02111} & 4.30~|~0.92~|~3.28  \\
& & & VALL-E Token~\citep{DBLP:journals/corr/abs-2301-02111} & \textbf{6.57}~|~\textbf{0.93}~|~\textbf{3.28} \\
& & & LauraGPT & 8.62~|~0.91~|~3.26 \\
\bottomrule
\end{tabular}
}
\end{table*}

\section{Results and Analysis}
Section~\ref{sec:experiments} presents the main results of performance comparison on the basic tasks 
from the state-of-the-art (SOTA) model, a \textbf{comparable} baseline, and our LauraGPT. Ablation studies in Section~\ref{sec:analysis} demonstrate the advantages of using continuous representations for audio inputs in LauraGPT by comparing to a counterpart with both discrete inputs and outputs (denoted \textbf{Discrete IO}), the superiority of our one-step codec vocoder, and effectiveness of multi-task finetuning. Further analyses include comparison with related unified Audio-and-Text LLMs~(Appendix~\ref{appendix:comparison-other-text-audio-model}), more analysis of multi-task fine-tuning on SER task~(Appendix~\ref{ser-impact-of-multi-task-finetuning}), comparing batch normalization with layer normalization in the audio encoder~(Appendix~\ref{sec:normlization}), and studying impact of initialization from pre-trained models~(Appendix~\ref{sec:init-gpt}).

\subsection{Results on All Tasks}
\label{sec:experiments}
Table~\ref{tab:overall_res} shows the results from the SOTA model, a comparable baseline, and our LauraGPT\footnote{Our results are from single runs due to the stability of the models and limited computational resources.}, in that order, on a variety of speech recognition, understanding, and generation benchmarks. 
The SOTA model yields the best results on each test set based on our literature review. The baseline for each task is chosen to facilitate fair comparison with LauraGPT: they are comparable to LauraGPT in model architecture or training data and are also common competitive baselines in the literature.
We cite the SOTA results to validate that LauraGPT consistently performs competitively on all the speech recognition, understanding, and generation tasks and the baselines are competitive. However, LauraGPT results cannot be fairly compared to the SOTA results. Specifically, QwenAudio achieves SOTA performance on most speech understanding tasks, but compared to LauraGPT, QwenAudio uses a much larger LLM ($\sim$7B VS. our 1.8B LLM), and uses the Whisper audio encoder trained on a large amount of ASR data while we use a Conformer encoder trained on much less data. Moreover, QwenAudio does not support speech generative tasks hence cannot handle SE and TTS tasks. Paraformer-large and UniverSLU achieve SOTA results on AISHELL-2 test-ios for Chinese ASR and on SLURP test for SLU; however, they only support single tasks and also train on more data than LauraGPT on the corresponding task. 
Appendix~\ref{appendix:comparison-other-text-audio-model} shows that LauraGPT greatly outperforms Whisper Large V2 on Chinese ASR test sets while the gap on English ASR test sets are primarily attributed to the much smaller English data used for training LauraGPT. For TTS, the SOTA VALL-E Phone outperforms baseline VALL-E Token\footnote{We re-implement two VALL-E models with 0.34B trainable parameters, both trained with the same data as LauraGPT. VALL-E Phone uses phonemes as the text input representation, while VALL-E Token uses WordPiece tokens from the text tokenizer.}, suggesting the importance of text representation for TTS.
Compared to both VALL-E models, LauraGPT achieves comparable speaker similarity (SECS) and speech quality (MOSNet). The degradation in content consistency (WER) from LauraGPT 
results from the generalization issue, since the training data is too limited for LauraGPT with 2B parameters. Overall, the results show that \textbf{LauraGPT consistently achieves comparable to superior performance than strong baselines on diverse speech tasks, demonstrating the general effectiveness of LauraGPT on speech recognition, understanding, and generative tasks}.

\begin{table*}[htb]
\setlength\tabcolsep{8pt}
\centering 
\caption{Comparison of Discrete IO models and LauraGPT on ASR, S2TT, and SE tasks for analysis of discrete VS. continuous representations for audio inputs. The best results on each test set are in bold.}
\label{tab:diif_input_token}
\scalebox{0.85}{
\begin{tabular}{l c c c c } 
\toprule 
\textbf{Task}    & \textbf{Dataset}  & \textbf{Metric}  & \textbf{Discrete IO}     & \textbf{LauraGPT}   \\
\midrule
\multirow{4}{*}{\textbf{ASR}}   & AISHELL-1 test          & CER~$\downarrow$ & 7.1 & \textbf{1.8}  \\
                                & AISHELL-2 test-ios      & CER~$\downarrow$ & 8.6 & \textbf{3.2}  \\
                                & LibriSpeech test-clean  & WER~$\downarrow$ & 9.1 & \textbf{4.4}  \\
                                & LibriSpeech test-other  & WER~$\downarrow$ & 24.0 & \textbf{7.7}  \\
\midrule
\multirow{2}{*}{\textbf{S2TT}}  & BSTC dev~(Zh$\rightarrow$EN)          & BLEU~$\uparrow$ & 5.1 & \textbf{17.8}  \\
                                & CoVOST2 test set~(En$\rightarrow$Zh)      & BLEU~$\uparrow$ & 5.0 & \textbf{38.5}  \\
\midrule
\multirow{3}{*}{\textbf{SE}} & \multirow{3}{*}{\makecell{Mixup of LibriSpeech\\test-clean, FSD50K and\\noise-92}}    & PESQ~$\uparrow$ & 1.96 & \textbf{2.97}  \\
                                             &       & STOI~$\uparrow$ & 64.0 & \textbf{88.0}  \\
                                             &       & WER~$\downarrow$ & 53.97 & \textbf{15.94}  \\
\bottomrule
\end{tabular}
}
\end{table*}

\vspace{-2mm}
\subsection{Analysis}
\vspace{-2mm}
\noindent \paragraph{Discrete VS. Continuous Representations for Audio Inputs}
\label{sec:analysis}
Existing unified Audio-and-Text LLMs use discrete tokens to represent audio inputs. We analyze the efficacy of using continuous representations for audio inputs in LauraGPT by comparing to its counterpart \textbf{Discrete IO} on ASR, S2TT, and SE tasks, representing \textbf{audio-input recognition and understanding, and audio generation capacities}. In Discrete IO,
both audio inputs and outputs are represented by flattened codec tokens from \textit{the first four quantizers}\footnote{Using outputs of the first quantizer (as in LauraGPT) for audio tokenizer renders very poor performance for audio-input tasks with the Discrete IO models. Using more quantizers improves performance but reduces efficiency.}, resulting in a token rate of 100Hz. In LauraGPT, audio inputs are represented by continuous acoustic features, which are also fed into our one-step vocoder as a condition to achieve high-quality outputs. Table~\ref{tab:diif_input_token} shows that \textbf{LauraGPT consistently outperforms Discrete IO with remarkable gains on all tasks}. \textbf{For ASR task}, the performance degrades drastically when replacing continuous features with discrete audio tokens. Although the performance degradation can be reduced by using more quantizers (more codec groups), e.g. 32~\citep{DBLP:journals/corr/abs-2309-10922}, more codec groups always cause higher token rates and longer sequence and in turn higher computational demands.
\textbf{For S2TT task}, Discrete IO only yields BLEU scores of 5.1 and 5.0 on test sets,  basically suggesting lack of translation capability.
\textbf{For SE task}, using codec tokens as inputs cannot improve the quality and intelligibility of noisy speeches, suggesting lack of enhancement capability, 
probably because the distribution of noisy speech is too complicated to be accurately represented by four groups of discrete audio tokens.

\begin{table}[t!]
\centering 
\caption{Comparison of our one-step audio synthesis scheme and the multi-step scheme on the SE task.}
\label{tab:comp-audio-syn-se}
\scalebox{0.83}{
\begin{tabular}{l c c c c } 
\toprule 
\textbf{Scheme}    & \textbf{PESQ}~$\uparrow$  & \textbf{STOI(\%)}~$\uparrow$  & \textbf{CER}~$\downarrow$    & \textbf{WER}~$\downarrow$   \\
\midrule
\textbf{Multi-step} & 2.55 & \textbf{88.0} & 10.52 & 19.32 \\
\textbf{One-step} & \textbf{2.97} & \textbf{88.0} & \textbf{9.05} & \textbf{15.94} \\
\bottomrule
\vspace{-1cm}
\end{tabular}
}
\end{table}

\noindent \paragraph{Comparison on Audio Synthesis Schemes}
\label{sec:comp-audio-syn}
VALL-E~\citep{DBLP:journals/corr/abs-2301-02111} introduces a commonly used scheme formulating audio synthesis as a \textit{classification problem}: A neural network is shared to predict the codec tokens in the following group with the previous ones as inputs and synthesizing target audio requires multiple steps or iterations to achieve a reasonable speech quality. In contrast, our one-step codec vocoder formulates audio synthesis as a \textit{regression problem}. As described in Section~\ref{sec:codec_vocoder}, our one-step codec vocoder simplifies audio synthesis into a single feed-forward calculation and overcomes the prediction challenge caused by the multimodal distribution of codec tokens. Table~\ref{tab:comp-audio-syn-se} 
shows that \textbf{our one-step codec vocoder greatly outperforms the multi-step scheme in terms of content consistency (CER, WER) and speech quality (PESQ), while obtaining the same intelligibility (STOI)}.

\noindent \paragraph{Effectiveness of Multi-task Finetuning}
\label{sec:impact-of-multi-task-finetuning}
The multi-task fine-tuned LauraGPT (Section~\ref{sec:task-details}) could be advantageous over individual single-task models: (1) Multi-task learning could exploit synergy between tasks and reduce over-fitting, in turn yield high performance on diverse tasks and achieve better performance than single-task training.  (2) Multi-task learning could learn a single model capable of supporting a wide range of tasks, hence practical deployment is greatly simplified through unified model implementation and API.


We investigate whether the multi-task trained LauraGPT could achieve better performance than single-task training for tasks with limited training data. Among the basic tasks (Table~\ref{tab:datasets}), AAC, SLU, and SER tasks all have limited training data. 
We initialize the Qwen backbone and the audio encoder the same as LauraGPT before conducting multi-task training, 
then train the single-task model only using the task-specific training data. The results are shown in Table~\ref{tab:single-multi-compare}.  

\textbf{For the AAC task}, we find that the multi-task trained LauraGPT outperforms the single-task model on SPICE, CIDEr and SPIDEr on the Clotho evaluation set. \textbf{For the SLU task}, on the SLURP test set, LauraGPT greatly outperforms the single-task model on intent accuracy by \textbf{+2.9} absolute and on SLU-F1 by \textbf{+22.5} absolute.  \textbf{For the SER task}, on the MELD test set, LauraGPT substantially outperforms the single-task model in terms of UA and the primary WF1 metrics, while the WA result is slightly worse. More analyses in Appendix~\ref{ser-impact-of-multi-task-finetuning} show that multi-task learning dramatically improves accuracies of the minority classes. \textbf{In summary, these results verify that multi-task learning for LauraGPT consistently achieves better performance than single-task training for tasks with limited training data.}

\begin{table}[htb]
\centering 
\caption{Comparison of single-task finetuning and multi-task finetuning on the AAC, SLU, and SER tasks.}
\label{tab:single-multi-compare}
\scalebox{0.85}{
\begin{tabular}{l c c c c } 
\toprule 
\textbf{Task}    & \textbf{Dataset}  & \textbf{Metric}  & \textbf{Single}     & \textbf{Multi}   \\
\midrule
\multirow{3}{*}{\textbf{AAC}}   & \multirow{3}{*}{Clotho eval} & SPICE~$\uparrow$      & 0.07  & \textbf{0.08} \\
&      & CIDEr~$\uparrow$     & 0.16   & \textbf{0.22} \\
                                                        &      & SPIDEr~$\uparrow$     & 0.11   & \textbf{0.15} \\
\midrule
\multirow{2}{*}{\textbf{SLU}}   & \multirow{2}{*}{SLURP test}  & Intent ACC~$\uparrow$ & 85.0 & \textbf{87.9} \\
                                                            &  & SLU-F1~$\uparrow$     & 51.0 & \textbf{73.5} \\
\midrule
\multirow{3}{*}{\textbf{SER}}   & \multirow{3}{*}{MELD test}   & WA~$\uparrow$   & \textbf{0.508} & 0.507 \\
                                                          &    & UA~$\uparrow$   & 0.221 & \textbf{0.312} \\
                                                          &    & WF1~$\uparrow$  & 0.426 & \textbf{0.492} \\
\bottomrule
\end{tabular}
}
\end{table}




\vspace{-4mm}
\section{Conclusion}
\label{sec:conclusion}
\vspace{-2mm}
We propose LauraGPT that can handle both audio and text inputs and outputs and perform audio recognition, understanding, and generation. We propose combining continuous and discrete features for audio and a one-step codec vocoder, and employ multi-task learning. Experiments demonstrate that LauraGPT achieves comparable to superior performance compared to strong baselines on a wide range of speech tasks on content, semantics, paralinguistics, and audio-signal analysis. 

\section*{Limitations}
In this work, in order to support a wide range of audio recognition, understanding, and generation tasks, we choose to train all parameters in LauraGPT during supervised multi-task finetuning, including the Qwen backbone, except for the codec vocoder. This strategy results in substantial computations for training. In future work, we plan to investigate parameter-efficient fine-tuning to reduce computation demands. Also, due to the limited computation resources, our comparisons between the multi-task trained LauraGPT and single-task models are focused on the low-resource tasks, that is, AAC, SLU, and SER tasks. We find that multi-task learning for LauraGPT consistently achieves better performance than single-task training for tasks with limited training data. Next, we plan to complete comparisons of LauraGPT and single-task models on all tasks, including relatively rich-resource tasks such as ASR. These studies will promote understandings on where tasks could benefit from each other, including tasks with even conflicting objectives. We also plan to conduct deeper analysis on the potential risk of catastrophic forgetting of the original text capabilities of the pre-trained text LLM, due to multi-task learning of speech tasks. Note that exploration of parameter-efficient fine-tuning may also help preserve the original text capabilities of the pre-trained text LLMs.

LauraGPT relies on discrete audio tokens for speech generative tasks. Our research shows that the performance of this paradigm strongly depends on the quality of the audio tokenizer. We plan to systematically analyze the impact of various audio tokenizers on diverse audio generative tasks. We plan to develop new audio tokenizers that are more suitable for unified Auio-and-Text LLMs and provide desirable representations for generative tasks.

There are great emerging interests in fundamental speech models that are similar to those in the field of NLP. This is a tremendously valuable research direction. Our work achieves  important milestone for this research question, as we explore and provide promising answers to the following question: \textit{How to design more efficient and scalable unified GPT-style Audio-and-Text LLMs than existing approaches that can leverage large-scale labeled data and achieve highly competitive performance on a diverse set of speech tasks, including speech recognition, understanding and generation, using a single model?} Note that previous general speech models either focus solely on speech recognition and understanding tasks but neglect speech generative tasks, or support speech generation but suffer from severe performance degradation on speech recognition and understanding tasks.

Inspired by the recent advances of LLMs in NLP, we envision that the fundamental speech models should have the following capabilities:

\begin{itemize}
    \item In-context learning ability like GPT-3, which can learn from few-shot examples and adapt to new tasks, such as predicting the age of the speaker from a speech sample.
    \item Instruction-following ability like InstructGPT and ChatGPT, which can perform the appropriate speech-related task given a natural language instruction, such as synthesizing a speech with a specific emotion or style.
    \item General audio modeling abilities, i.e., speech, non-speech audio, and music, such as music generation.
\end{itemize}

Our work demonstrates that the current LauraGPT has made solid progress and reached one important milestone toward a speech foundation model. 
From LauraGPT to the next-generation speech foundation model we envision, most remaining efforts are in more task data collection and more self-supervised and/or supervised pre-training and supervised fine-tuning. There is no need to modify the model architecture.

\bibliography{custom}

\clearpage
\appendix
\appendixpage

\section{Experimental Details}
\label{appendix:details}
\subsection{Basic Tasks}
\label{sec:task-intro}
The following tasks are used in supervised multi-task learning of LauraGPT and also in evaluations:

\noindent \textbf{Automatic speech recognition (ASR)} is a vital task in the speech processing community. It focuses on transcribing speech into textual content. 

\noindent \textbf{Spoken language understanding (SLU)} is a task of directly deriving high-level semantic meaning from audio input. It aims to identify the user’s intent and the relevant entity slots that fill the intent. An intent is usually composed of a scenario type and an action type, while slots and fillers are key-value pairs that specify the details of the intent.

\noindent  \textbf{Speech-to-text translation (S2TT)} is similar to machine translation, but it directly translates the source language speech into the target language text. 

\noindent \textbf{Speech emotion recognition (SER)} categorizes the emotions in speech input. Compared to textual emotion recognition, speech signals convey additional information, including tone and speaking rate, which enhances emotion recognition.

\noindent \textbf{Automated audio captioning (AAC)} aims to generate a natural language sentence that describes the content of an audio clip.

\noindent \textbf{Speech enhancement (SE)} is an audio-to-audio task that aims to improve speech quality through noise suppression and dereverberation. 
In order to incorporate this task into a unified modeling framework, we reformulate the task as a classification problem using codec tokens.

\noindent \textbf{Text-to-speech synthesis (TTS)} can be considered as the inverse process of ASR, where it generates speech that matches the given text.


    
\subsection{Training Datasets}
\label{sec:training-datasets}
To ensure reproducibility, all training data and test data for LauraGPT are publicly available datasets, with licenses of Apache 2.0, CC BY 4.0, CC0, non-commercial research and education use, etc.
The training data for the basic tasks listed in Section~\ref{sec:task-details} and defined in Appendix~\ref{sec:task-intro} are prepared as follows. 

For the ASR task, we utilize open-source Chinese datasets such as AISHELL-1~\citep{dataset_aishell1},
AISHELL-2~\citep{dataset_aishell2},
WenetSpeech~\citep{dataset_wenetspeech}, 
as well as open-source English datasets including LibriSpeech~\citep{dataset_librispeech} 
and GigaSpeech~\citep{dataset_gigaspeech}.

For the S2TT task, we employ the commonly used BSTC~\citep{dataset_bstc} and CoVOST 2~\citep{dataset_covost_2} 
datasets. Due to the limited data volumes of BSTC and CoVoST 2, we further augment the training set by translating AISHELL-1 and AISHELL-2 datasets into English and translating LibriSpeech dataset into Chinese using a publicly available text translation model~\citep{wei-etal-2022-learning}.
Consequently, we obtain approximately 2,000 hours of supplementary data for Chinese-to-English and English-to-Chinese S2TT tasks. As a supplement of training data for S2TT, we also add the ParaCrawl v9 dataset~\citep{dataset_mt}, which consists of 14M parallel text sentences for Zh$\rightarrow$En (Chinese-to-English) and En$\rightarrow$Zh (English-to-Chinese) translations.

For the SER task, we collect corpora including MELD~\citep{dataset_meld}, IEMOCAP~\citep{dataset_iemocap},  RAVDESS~\citep{dataset_ravdess}, TESS~\citep{dataset_tess}, Crema-D~\citep{dataset_crema}, Emov-DB~\citep{dataset_emov-db}, and SAVEE~\citep{dataset_savee}. These corpora are recorded in multi-modal formats, comprising audio or visual data. No other corpora are used for the SER task. 

For the SLU task, we use the multi-domain Spoken Language Understanding Resource Package~(SLURP) dataset~\citep{dataset_slurp}, which covers 18 scenarios.

For the AAC task, we use AudioCaps~\citep{dataset_audiocaps}, WavCaps~\citep{dataset_wavcaps}, and Clotho~\citep{dataset_clotho} datasets. 

For the SE task,  pairs of noisy and clean speech are required for training. 
The clean utterances are extracted from the AISHELL-1, AISHELL-2, LibriSpeech, and WSJ datasets~\cite{DBLP:conf/interspeech/PaulB92}, while the noisy counterparts are generated by mixing the clean speech with noises from the FSD-50K dataset~\citep{DBLP:journals/taslp/FonsecaFPFS22} at random signal-to-noise rates (SNR) ranging from 2 to 15. Besides the additional noises, we also simulate convolutional noises by convolving the clean speech data with room impulse responses~\citep{ko2017study}. As a result, we obtain approximately 6000 hours of paired data for the SE task.


For the TTS task, we use the open-source LibriTTS and 3D-speaker datasets~\citep{zheng20233d}.
Further details of the training data for all tasks can be found in Table~\ref{tab:datasets}.


Note that for all the training and test datasets, our use of the data is consistent with their intended use. We use all data sets in the same ways as prior research works, hence we did not check whether the data that was used contains any information that names or uniquely identifies individual people or offensive content.

\begin{table*}[htb]
\renewcommand{\arraystretch}{1.5}
\renewcommand\cellgape{\Gape[3pt]}
\setcellgapes{5pt}
\centering 
\caption{Statistics of the training data for basic tasks in Section~\ref{sec:task-details}. $\text{Corpus}^{\times N}$ means that the training samples in this corpus are copied $N$ times during training.} 
\label{tab:datasets}
\vspace{0.1cm}
\begin{tabular}{l  l  c } 
\toprule 
\textbf{Task} & \textbf{Training Data} & \textbf{\# Samples} \\
\midrule
\textbf{ASR}  & \makecell[l]{AISHELL-1, AISHELL-2, WenetSpeech,  LibriSpeech, GigaSpeech} &  24.2 M\\
\textbf{SLU}  & SLURP$^{\times 10}$   & 1.2 M  \\
\textbf{S2TT} & \makecell[l]{BSTC$^{\times 5}$, CoVOST 2$^{\times 2}$, AISHELL-1, 
AISHELL-2, LibriSpeech}   & 2.2 M         \\
\textbf{SER}  & \makecell[l]{MELD$^{\times 10}$, IEMOCAP$^{\times 10}$, RAVDESS$^{\times 10}$, TESS$^{\times 10}$\\ Crema-D$^{\times 10}$, Emov-DB$^{\times 10}$, SAVEE$^{\times 10}$}   & 0.3 M \\
\textbf{AAC}  & Clotho$^{\times 10}$, AudioCaps$^{\times 10}$, WavCaps$^{\times 5}$   & 1.3 M \\
\textbf{SE}   & \makecell[l]{AISHELL-1$^{\times 3}$, AISHELL-2$^{\times 3}$, LibriSpeech$^{\times 3}$, WSJ$^{\times 2}$, FSD-50K$^{\times 2}$, RIR} & 5.3 M\\
\textbf{TTS}  & LibriTTS$^{\times 2}$, 3D-Speaker$^{\times 2}$, AISHELL-1$^{\times 2}$, AISHELL-2$^{\times 2}$, LibriSpeech$^{\times 2}$ &  5.0 M  \\
\bottomrule
\end{tabular}
\end{table*}

\subsection{Evaluation Datasets and Metrics}
\label{sec:evaldata-metrics}
Table~\ref{tab:eval-data-metric} presents the evaluation datasets and evaluation metrics for various tasks. The metrics used in our experiments are described below:
\begin{itemize}[leftmargin=*,noitemsep]
    \item \textbf{CER} stands for Character Error Rate, a commonly used metric to evaluate the recognition performance of Chinese and English utterances. We also utilize CER to assess the content consistency in TTS task.
    \item \textbf{WER} stands for Word Error Rate, which considers entire words rather than individual characters. In our experiments, we use WER to evaluate ASR recognition performance, content consistency in TTS, and speech intelligibility in SE.
    \item \textbf{SECS}, which stands for Speaker Encoder Cosine Similarity, utilizes speaker embeddings extracted from a pre-trained speaker verification model \footnote{Code is available at \url{https://huggingface.co/microsoft/wavlm-base-plus-sv}} for both prompt and synthesized speech. The cosine similarity between the two embeddings is then employed to measure the speaker similarity between the prompt speech and the synthesized speech. Furthermore, the naturalness of the synthesized speech is evaluated using \textbf{MOSNet}, a non-intrusive score derived from a pre-trained neural network \footnote{Code is available at \url{https://github.com/lochenchou/MOSNet}}.
    \item \textbf{BLEU} represent the Bilingual Evaluation Understudy metric. BLEU is commonly used to assess the quality of machine-generated text by comparing it to reference translations. In our experiments, we use BLEU to evaluate S2TT.
    \item \textbf{PESQ} represents Perceptual Evaluation of Speech Quality, while \textbf{STOI} stands for Short-time Objective Intelligibility. Both metrics are widely used to assess speech enhancement. PESQ ranges from $-0.5$ to $4.5$, whereas STOI is in the range of $[0,1]$.
    \item \textbf{SPICE}, \textbf{CIDEr} and \textbf{SPIDEr} are metrics borrowed from the image captioning task and employed for AAC evaluation. SPICE stands for Semantic Propositional Image Caption Evaluation, CIDEr denotes Consensus-based Image Description Evaluation, and SPIDEr represents the average of SPICE and CIDEr.
    \item \textbf{WA}, \textbf{UA} and \textbf{WF1} stands for weighted accuracy, unweighted accuracy and the weighted F1 score. WA corresponds to the overall accuracy, UA corresponds to the average class-wise accuracy, and WF1 corresponds to the average class-wise F1 score. 
    \item \textbf{ACC} measures the accuracy of predicting the intent. \textbf{SLU-F1} is a metric that balances Word-F1 and Char-F1, computed as the sum of the confusion matrices.
\end{itemize}


\begin{table*}[htb] 
\renewcommand{\arraystretch}{1.5}
\renewcommand\cellgape{\Gape[3pt]}
\centering 
\caption{Evaluation datasets and metrics for different tasks. $\uparrow$ indicates that higher values of the metric are desirable, while $\downarrow$ implies the opposite.} 
\label{tab:eval-data-metric}
\scalebox{0.85}{
\begin{tabular}{l l l } 
\toprule
\textbf{Task} & \textbf{Evaluation Datasets} & \textbf{Evaluation Metrics} \\
\midrule
\textbf{ASR} & \makecell[l]{AISHELL-1 test, AISHELL-2 test-ios, \\ Librispeech test-clean \& test-other} & CER~$\downarrow$, WER~$\downarrow$ \\
\textbf{SLU} & SLURP test & ACC~$\uparrow$, SLU-F1~$\uparrow$ \\
\textbf{S2TT} & BSTC dev, En$\rightarrow$Zh subset of CoVOST2 & BLEU~$\uparrow$ \\
\textbf{SER} & MELD test & WA~$\uparrow$, UA~$\uparrow$, WF1~$\uparrow$ \\
\textbf{AAC} & Clotho eval &  SPICE~$\uparrow$,
CIDEr~$\uparrow$,
SPIDEr~$\uparrow$\\
\textbf{SE} & LibriSpeech test-clean, FSD50K, noise-92 & PESQ~$\uparrow$, STOI~$\uparrow$, WER~$\downarrow$ \\
\textbf{TTS} & AISHELL-1 test, LibriTTS test-clean & CER~$\downarrow$, WER~$\downarrow$, SECS~$\uparrow$, MOS~$\uparrow$ \\
\bottomrule
\end{tabular}
}
\end{table*}

\subsection{Details of Training and Inference}
\label{sec:detail_training_setup}

In all experiments, we optimize the model parameters through the following steps: (1) We initialize the Qwen backbone and the audio encoder with the pre-trained checkpoints. (2) We then perform multi-task finetuning.

Due to the significant variation in data volume across different tasks, the training process is conducted in three stages.
In the first training stage, the model is fine-tuned on all tasks using the complete training data as shown in Table~\ref{tab:datasets}. The AdamW optimizer is utilized with a peak learning rate of $5 \times 10^{-4}$ and 10K warmup steps. In the second stage, we further fine-tune the model on tasks that have small-scale datasets, including TTS, SE, AAC, SER, and SLU tasks. The AdamW optimizer is utilized with a peak learning rate of $2 \times 10^{-4}$ and 10K warmup steps. 
In the third training stage, we fine-tune the model on all tasks using the complete training data again. The peak learning rate of the AdamW optimizer for the third stage is reduced by half as $1\times 10^{-4}$, while the warmup step remains at 10K.

For the codec vocoder, we train the predictor on the training data of the TTS and SE tasks. We use the Adam optimizer with a peak learning rate of 0.001 and 25K warmup steps. 
The decoder of the codec vocoder is initialized with the pre-trained checkpoints\footnote{\url{https://funcodec.github.io}} and kept frozen during the multi-task finetuning of LauraGPT.

As stated in Section~\ref{sec:methods}, during the training stage, the input is converted into input embeddings by the audio encoder if the input is audio, or converted by the embedding matrix $W$ if the input is text, while the output is converted into output embeddings by the same embedding matrix $W$ for teacher-forcing. Meanwhile, this matrix $W$ is also used to convert the task-ID token into an embedding. Then, these embeddings are composed into an embedding sequence as [input embeddings, task-ID embedding, output embeddings], which is taken as the input of Qwen LLM. To train the model, a masked cross-entropy loss function is applied, as shown in Eq.~\ref{eq:LM_loss}. As described in Section~\ref{sec:methods}, in addition to masking out the losses on inputs, the cross-entropy loss at the position of the task token is also masked out.

During the inference stage, the input is converted into input embeddings as done during the training stage. Then the corresponding task-ID embedding is added at the end of the input embedding sequence. Next, the Qwen LLM generates output tokens in an autoregressive manner until the ``end of sequence'' token is generated. Finally, for text-format output, the Qwen tokenizer is employed to convert tokens into final output, while for audio-format output, the codec vocoder is employed to convert tokens into raw waveforms.

\subsection{Details of the SER Evaluation}
\label{sec:detail_ser}
During the training stage, emotion labels within different training corpora are unified into the following nine classes: anger, disgust, neutral, like, sadness, surprise, happiness, joy, and fear.
At the test stage, we map the ``like'' and ``happiness'' emotion classes into the ``joy'' class to match the MELD test set. 
LauraGPT uses an autoregressive structure to generate emotion labels. Out-of-domain outputs are considered as classification errors, making the task harder. 
Both WavLM Base model and WavLM Large model utilize the weighted sum of multiple layers with learnable parameters as speech features, which are fed into a downstream network for classification.

\section{Comparison with Related Unified Audio-and-Text Models}
\label{appendix:comparison-other-text-audio-model}

Table~\ref{tab:compare} compares our LauraGPT against the most related works, which, similar to LauraGPT, are all multi-task unified audio-and-text models. Due to the drastic differences in experimental settings, datasets used and lack of open source codebase and checkpoints, it is difficult to conduct a fair comparison between LauraGPT and these most related multi-task unified audio-and-text models. Despite all these difficulties, below we provide the most relevant results for comparing LauraGPT and these related models.

\begin{table*}[htb] 
\centering 
\vspace{-0.2cm}
\caption{Comparisons with the most related multi-task unified audio-and-text models. The table shows the tasks that each model is trained and evaluated on.}
\label{tab:compare} 
\scalebox{0.9}{
\begin{tabular}{l c c c c c  } 
\toprule 
 & \textbf{SpeechT5} & \textbf{Whisper} & \textbf{VioLA} &  \textbf{AudioPaLM} & \textbf{LauraGPT(Ours)}  \\
\midrule
\textbf{Date} &  2021.10 & 2022.12  & 2023.5 & 2023.6 & 2023.9 \\  
\textbf{Organization} & Microsoft & OpenAI  & Microsoft & Google & Ours \\
\textbf{Model Size} & 0.14B & 1.5B   & 0.25B & 8B & 2.0B \\
\textbf{Pair Data (hrs)} & 0.96K & 680K   & 79K & 48K   & 60K\\
\textbf{Unsup. Pretrain} & N/A & N/A    & N/A & PaLM-2  & Qwen-1.8B \\
\textbf{Audio Input} & Continuous  & Continuous   & Discrete & Discrete & Continuous \\
\textbf{Audio Output} & N/A  & N/A   & Discrete & Discrete & Discrete  \\
\textbf{Languages} & EN & 99  &  EN/CN & 113 & EN/CN   \\
\midrule
\textbf{ASR} & \CHECK & \CHECK  & \CHECK & \CHECK &  \CHECK\\
\textbf{S2TT} & \CHECK & \CHECK  & \CHECK & \CHECK & \CHECK \\
\textbf{TTS} & \CHECK & \CROSS   & \CHECK & \CHECK &  \CHECK \\
\textbf{SE} & \CHECK & \CROSS    & \CROSS & \CROSS &  \CHECK\\
\textbf{AAC} & \CROSS & \CROSS    & \CROSS & \CROSS &  \CHECK\\
\textbf{SER} & \CROSS & \CROSS    & \CROSS & \CROSS &  \CHECK\\
\textbf{SLU} & \CROSS & \CROSS    & \CROSS & \CROSS &  \CHECK\\
\bottomrule
\end{tabular}
}
\end{table*}

\textbf{Whisper}~\citep{DBLP:journals/corr/abs-2212-04356} is solely studied on the ASR task in the original paper, hence we compare LauraGPT to Whisper only on the ASR task. As shown in Table~\ref{tab:asr_result}, on the Chinese test sets AISHELL-1 test and AISHELL-2 test-ios, LauraGPT greatly outperforms Whisper by \textbf{-3.9} and \textbf{-2.3} absolute on CER with much smaller training data. On the English test sets Librispeech test-clean and test-other, LauraGPT performs worse than Whisper Large V2 as Whisper Large V2 uses much more English training data than LauraGPT.

\begin{table*}[htb] 
\centering 
\vspace{-0.2cm}
\caption{Comparison of different models on the ASR task in terms of CER(\%) $\downarrow$ for Chinese and WER(\%) $\downarrow$ for English. Data size denotes the number of hours.} 
\label{tab:asr_result}
\vspace{0.1cm}
\setlength\tabcolsep{4pt}
\begin{tabular}{l |c| c | c c c c} 
\toprule 
\textbf{Model} & \textbf{\makecell[c]{Model \\ Size}}    & \textbf{\makecell[c]{Data \\ Size}} & \textbf{\makecell[c]{AISHELL-1 \\ test}}   & \textbf{\makecell[c]{AISHELL-2 \\ test-ios}} & \textbf{\makecell[c]{Librispeech \\ test-clean}}  & \textbf{\makecell[c]{Librispeech \\ test-other}} \\
\midrule
\textbf{Paraformer~(CN)}  & 0.2 B & 60K         & 2.0      & 2.9      & -         & -  \\
\textbf{Paraformer~(EN)}  & 0.2 B & 20K         & -      & -     & 3.5         & 8.2  \\
\textbf{Whisper Large V2} & 1.5 B  & 680K         &  5.7         & 5.5         & 2.5      & 4.9 \\
\midrule
\textbf{LauraGPT (Ours)}       & 1.8 B  & 22K         & 1.8      & 3.2      & 4.4      & 7.7  \\
\bottomrule
\end{tabular}
\end{table*}

\textbf{SpeechT5}~\citep{DBLP:conf/acl/AoWZ0RW0KLZWQ0W22} is evaluated on ASR, TTS, S2TT, voice conversion (VC), SE, and speaker identification (SID). Since the training data of tasks other than ASR for SpeechT5 differs remarkably from those for LauraGPT, we compare LauraGPT against SpeechT5 only on ASR. For SpeechT5, the model is first pre-trained with large-scale unlabeled speech and text data. Then, it is finetuned on the Librispeech-960 corpus via the hybrid cross-entropy and CTC loss. As claimed in their paper, SpeechT5 achieves a WER of 7.3\% on the Librispeech test-other subset without CTC and LM. Under a fair comparison, our LauraGPT achieves a comparable WER of 7.7\%. \textbf{Note that different from SpeechT5, LauraGPT is directly trained on multi-task labeled datasets without benefiting from any self-supervised pre-training}.

\textbf{VioLA}~\citep{DBLP:journals/corr/abs-2305-16107} is evaluated on ASR, S2TT, TTS and S2ST tasks. Considering the substantial differences in training data on tasks between VioLA and LauraGPT and lack of open-sourced VioLA codebase and models, it is difficult to fairly compare LauraGPT with VioLA. Among the tasks, direct comparison on ASR is also challenging since VioLA only conducts speech-to-phoneme recognition and reports Phoneme Error Rate (PER) rather than recognizing words/characters and reporting WER/CER as conducted by LauraGPT. According to their paper, VioLA underperforms their in-house Attention-based Encoder-Decoder (AED) model (which we also have no access to) with relative 19.96\% phoneme error rate (PER) degradation from 9.47\% to 11.36\% on Mandarin WenetSpeech dev set. Since higher PER always corresponds to much higher WER as a word comprises multiple phonemes, it would be safe to hypothesize that the relative degradation on WER from VioLA over AED is even greater. In contrast, compared with the Paraformer baseline, our LauraGPT achieves comparable CER on the Mandarin AISHELL-2 test-ios set and outperforms it on the English Librispeech test-other set, i.e., overall LauraGPT performs comparably to Paraformer. Note that Paraformer is a non-autoregressive AED model performing comparably to conventional auto-regressive AED model~\citep{gao2022paraformer}. Therefore, \textbf{through this chain of comparisons, we are confident to conclude that LauraGPT notably outperforms VioLA on ASR task}.

\textbf{AudioPaLM}~\cite{DBLP:journals/corr/abs-2306-12925} is evaluated on ASR, S2TT and TTS tasks. Since the training and evaluation datasets for AudioPaLM and LauraGPT are disjoint, their performance results cannot be directly compared. In addition, the pre-trained model of AudioPaLM has not been released. Therefore, empirically comparing LauraGPT to AudioPaLM will require great effort and is not conducted in this work.

\begin{table}[htb] 
\centering 
\caption{Comparison of batch normalization~(BN) and layer normalization~(LN) on the SE task in terms of Loop Ratio (\%), PESQ and STOI(\%). $\uparrow$ indicates that higher values are desired, while $\downarrow$ implies the opposite.}
\label{tab:bn-ln} 
\scalebox{0.84}{
\begin{tabular}{l c c c} 
\toprule 
\textbf{Norm}  & \textbf{Loop Ratio~$\downarrow$}  & \textbf{PESQ~$\uparrow$}     & \textbf{STOI~$\uparrow$}  \\
\midrule
\textbf{BN} & 86.00 & 1.27  & 22.0 \\
\textbf{LN} & 4.60 & 2.97  & 88.0 \\
\bottomrule
\end{tabular}
}
\end{table}  


\section{More Analyses of Critical Design Choices}
\label{sec:further-analysis}

\subsection{Effectiveness of Multi-task Finetuning on the SER task}
\label{ser-impact-of-multi-task-finetuning}

Table~\ref{tab:single-multi-compare} shows that for the SER task, on the MELD test set, the multi-task trained LauraGPT substantially outperforms the single-task model in terms of UA and WF1 metrics, while the WA result is slightly worse. 

To further analyze the results of the SER task, we conduct a statistical analysis of the number of samples for each emotion class in both training and test sets of the MELD dataset, as well as their corresponding test accuracy. The results are shown in Table~\ref{tab:ser_result_class_compare}. Compared to the single-task model, the multi-task trained LauraGPT results in degradation in accuracy for classes with a larger number of training samples, while greatly improving the accuracy on classes with fewer training samples. This explains why WA decreases slightly from multi-task training while UA and WF1 show remarkable improvements. Note that \textbf{WF1 is the primary metric on the MELD dataset due to sample imbalance across different emotion classes~\cite{chen2023vesper}}. That is, on the primary metric WF1, the multi-task trained LauraGPT greatly outperforms the single-task model.  Furthermore, the accuracy of the \textit{disgust} and \textit{fear} classes from the single-task model is 0, which aligns with the fact that these two classes have the fewest training samples in the MELD dataset.  Multi-task training not only remarkably improves the performance of emotion classes with low accuracy (\textit{joy}, \textit{sadness}, \textit{surprise}), but also greatly improves the performance of classes that cannot be predicted with single-task training (\textit{disgust}, \textit{fear}). 


\begin{table*}[htb] 
\centering 
\caption{Accuracy on different emotion classes in the SER task from single-task finetuning and multi-task finetuning.}
\label{tab:ser_result_class_compare} 
\scalebox{0.84}{
\begin{tabular}{l c c c c c c c} 
\toprule 
\textbf{Model}          & \textbf{anger}   & \textbf{disgust}   & \textbf{neutral} & \textbf{joy} & \textbf{sadness} & \textbf{surprise} & \textbf{fear} \\
\midrule
\textbf{\#Training Samples}     & 1109  & 271 & 4710 & 1743  & 683 & 1205 & 268 \\
\textbf{\#Testing Samples}      & 345  & 68 & 1256 & 402  & 208 & 281 & 50 \\
\textbf{Single-task}            & 0.396  & 0.000  & 0.875  & 0.119  & 0.029  & 0.128  & 0.000 \\
\textbf{LauraGPT}               & 0.333  & 0.103  & 0.708  & 0.381  & 0.236  & 0.381  & 0.040 \\
\bottomrule
\end{tabular}
}
\end{table*}

\subsection{Batch normalization versus layer normalization in audio encoder}
\label{sec:normlization}
In the original design, batch normalization is applied after the convolution module in the Conformer-based audio encoder. 
However, we discover that this choice leads to endless looping decoding due to inaccurate estimations of mean and variance, particularly for tasks with long sequence lengths. When the issue of endless looping decoding occurs, the model generates several fixed tokens repeatedly and cannot stop the generation until achieving a pre-defined maximum length.
To address this issue, we replace batch normalization with layer normalization, which is more robust to various mini-batch sizes. We validate this design by focusing on the SE task, which generally has the longest sequence among all the included tasks. The results are shown in Table~\ref{tab:bn-ln}. BN means batch normalization while LN means layer normalization. To evaluate the occurring probability of endless loop decoding, we define the metric, ``loop ratio'', which represents the fraction of endless decoded cases among all test cases.
The results indicate that batch normalization causes a significantly high loop ratio at the inference stage, leading to unacceptable PESQ and STOI scores. In contrast, \textbf{by replacing batch normalization with layer normalization, we observe a considerable reduction in the loop ratio to a very low level, thereby greatly improving the speech enhancement performance}. It should be noted that although the loop ratio of layer normalization is restricted, further research is still desired to explore more general normalization methods suitable for all audio-and-text tasks.

\begin{table*}[htb]
\centering 
\caption{Impact of initialization on the ASR, S2TT and SE tasks.}
\label{tab:asr-s2tt-se-init}
\scalebox{0.853}{
\begin{tabular}{l c c c c } 
\toprule 
\textbf{Task}    & \textbf{Dataset}  & \textbf{Metric}  & \textbf{w/o init}   & \textbf{LauraGPT}   \\
\midrule
\multirow{4}{*}{\textbf{ASR}}   & AISHELL-1 test          & CER~$\downarrow$ & 4.3  & \textbf{1.8}  \\
                                & AISHELL-2 test-ios      & CER~$\downarrow$ & 6.0  & \textbf{3.2}  \\
                                & LibriSpeech test-clean  & WER~$\downarrow$ & 8.3  & \textbf{4.4}  \\
                                & LibriSpeech test-other  & WER~$\downarrow$ & 17.6 & \textbf{7.7}  \\
\midrule
\multirow{2}{*}{\textbf{S2TT}}  & BSTC dev~(Zh$\rightarrow$En)          & BLEU~$\uparrow$ & 8.4 & \textbf{17.8}  \\
                                & CoVOST2 test set~(En$\rightarrow$Zh)      & BLEU~$\uparrow$ & 12.2 & \textbf{38.5}  \\
\midrule                           
\multirow{3}{*}{\textbf{SE}} & \multirow{3}{*}{\makecell{Mixup of LibriSpeech\\test-clean, FSD50K and\\noise-92}}    & PESQ~$\uparrow$ & 2.88 & \textbf{2.97}  \\
                                             &       & STOI~$\uparrow$ & 85.3 & \textbf{88.0}  \\
                                             &       & Loop Ratio~$\downarrow$ & 6.00 & \textbf{4.60}  \\
\bottomrule
\end{tabular}
}
\end{table*}

\subsection{Impact of initialization from pre-trained models}
\label{sec:init-gpt}
In LauraGPT, both the GPT backbone and audio encoder are initialized with the weights of pre-trained checkpoints.
We investigate how the initialization affects the performance of LauraGPT.
The experimental results for the ASR, S2TT and SE tasks are presented in Table~\ref{tab:asr-s2tt-se-init}.
From the results, we observe that the initialization has a significant impact on the performance of ASR and S2TT tasks, while its influence on the SE task is relatively limited.
This suggests that the prior knowledge learned by the GPT backbone is crucial for text generation tasks, but less important for audio generation tasks.
Consequently, we hypothesize that \textbf{a reasonable approach to enhance the quality of generated audios could be to pre-train the GPT backbone not only with text sequences but also with audio token sequences}.




\section{Supporting More Complex Tasks}
\label{appendix:more-complex-tasks}
As stated in Section~\ref{sec:task-combination}, with its modular and flexible design, LauraGPT provides an extensible framework to support complex tasks. By breaking a task into sub-tasks among the basic tasks used in training and cascading the raw inputs and model outputs of sub-tasks, LauraGPT can perform more complex tasks than the basic tasks.

Similar to the speech-to-speech translation (S2ST) example, LauraGPT can perform more complex tasks by chaining together basic tasks as described above. Here are a few examples of other complex tasks that LauraGPT can support rather than doing them one by one.

\noindent \paragraph{Rich transcription} We can extend LauraGPT to simultaneously transcribe audio into content, speaker information (speaker identification, etc), paralinguistic information (emotion, etc.) and high-level semantic information (intent, slots, etc.) by including different task IDs at the generation process. This approach could avoid error accumulation in a pipelined approach and is more efficient than performing these tasks individually.


\noindent \paragraph{Noise-robust ASR} We can implement noise-robust ASR by chaining tasks and creating the following input sequence: [noisy speech embedding, <SE>, embedding of the enhanced speech, <ASR>]. Since SE and ASR are jointly trained for LauraGPT, LauraGPT could effectively exploit embeddings of the original noisy speech and enhanced speech for noise-robust ASR.

\end{document}